\documentclass[a4paper,11pt]{article}
\usepackage{pos}

\title{Search for Baryon/Lepton number violation processes at BESIII}

\author*[a]{Xudong Yu}
\onbehalf{On behalf of the BESIII Collaboration}

\affiliation[a]{School of Physics,Peking University,\\
  209 Chengfu Road, Beijing, People's Republic of China}

\emailAdd{yuxd@stu.pku.edu.cn}

\abstract{The observed matter-antimatter asymmetry in the universe is a serious challenge to our understanding of nature. Baryon/lepton number violation (BNV/LNV) decays have been searched for in many experiments to understand this large-scale observed fact. We present the recent results from the BESIII experiment, including a search for BNV through $\Lambda-\bar{\Lambda}$ oscillation in the decays $J/\psi\to pK^-\bar{\Lambda}$ and $J/\psi\to\Lambda\bar{\Lambda}$. We also present searches for LNV in $D_s^+\to h^+h^0e^+e^-$ and $\omega/\phi\to\pi^+\pi^-e^+e^-$ decays, alongside probes into processes violating both baryon and lepton numbers simultaneously, such as $\Xi^0\to K^+e^-/K^-e^+$.}

\FullConference{The European Physical Society Conference on High Energy Physics (EPS-HEP2025)\\
7-11 July 2025\\
Marseille, France\\}

\begin{document}
\maketitle

\section{Introduction}
In the Standard Model (SM), the lepton number is conserved in interactions, and neutrinos are massless. However, the observation of neutrino oscillations provides strong evidence that neutrinos have non-zero mass. To explain this, theorists proposed the "see-saw" mechanism~\cite{Georgi:1974sy}, which attributes the small mass of the observed light neutrinos to the existence of a heavy Majorana neutrino. The Majorana neutrino can be manifested through the LNV decays by a change in the lepton number of the system by two units ($\Delta L=2$). Experimentally, $\Delta L=2$ processes can be searched for in hadron decays~\cite{Milanes:2018aku,Li:2024moj}.

The observed matter-antimatter asymmetry in the universe remains a major unsolved problem in physics. As originally pointed out by Sakharov, BNV is one of the three essential conditions for generating this asymmetry~\cite{Sakharov:1967dj}, motivating extensive experimental searches. In the Grand Unified Theories (GUT) and SM extensions, BNV processes are permitted if the baryon and lepton numbers change by the same amount, satisfying $\Delta (B-L)=0$~\cite{Georgi:1974sy}. Furthermore, other SM extensions incorporating next-to-leading order BNV effects via dimension-seven operators allow for $\Delta(B-L)=2$ processes~\cite{Babu:2012iv}.

The BESIII experiment operates in the $\tau$-charm energy region, with a center-of-mass range from 1.84 to 4.95 GeV~\cite{BESIII:2009fln,Li:2024pox,Song:2025pnt}. With a data set comprising 10 billion $J/\psi$ events and 20 fb$^{-1}$ $\psi(3770)$ data~\cite{Liao:2025lth}, BESIII provides a unique platform for searching for both BNV and LNV processes.

\section{Search for BNV processes}
\subsection{$\Lambda-\bar{\Lambda}$ oscillations}
A crucial test of BNV is the search for neutron-antineutron ($n-\bar{n}$) oscillation~\cite{Mohapatra:1980qe, Kuzmin:1970nx, Baldo-Ceolin:1994hzw}. If such oscillations exist, analogous processes like $\Lambda-\bar{\Lambda}$ may also expected. The experimental study of $\Lambda-\bar{\Lambda}$ oscillation serves as an independent probe of BNV and can provide significant insights into these processes. It has been proposed to search for this phenomenon at BESIII using $\Lambda$/$\bar{\Lambda}$ baryons produced in $J/\psi$ decays~\cite{Kang:2009xt}. The time-dependent oscillation probability for a beam of free $\bar{\Lambda}$ baryons to transition into a $\Lambda$ is given by:
\begin{equation}
    \mathcal{P}(\Lambda,t) = \sin^2(\delta m_{\Lambda\bar{\Lambda}}\cdot t)\cdot e^{-t/\tau_{\Lambda}},
\end{equation}
where $\delta m_{{\Lambda\bar{\Lambda}}}$ is the mass difference (or oscillation parameter), $t$ is the observation time, and $\tau_{\Lambda}$ is the $\Lambda$ lifetime. At BESIII, the experimentally measurable qauntity is the time-integrated oscillation rate:
\begin{equation}
    \mathcal{P}(\Lambda) = 
    \frac{\int_{0}^{\infty}\sin^2(\delta m_{\Lambda\bar{\Lambda}}\cdot t)\cdot e^{-t/\tau_{\Lambda}}\cdot dt}{\int_{0}^{\infty}e^{t/\tau_{\Lambda}}\cdot dt}.
\end{equation}
From this, the oscillation parameter can be derived as:
\begin{equation}
    (\delta m_{{\Lambda\bar{\Lambda}}})^{2} = \frac{\mathcal{P}(\Lambda)}{2\tau_{\Lambda}^2}.
\end{equation}

\subsection{Search for $\Lambda-\bar{\Lambda}$ oscillations via $J/\psi\to pK^-\bar{\Lambda}$}
\begin{figure}
    \centering
    \includegraphics[width=0.8\linewidth]{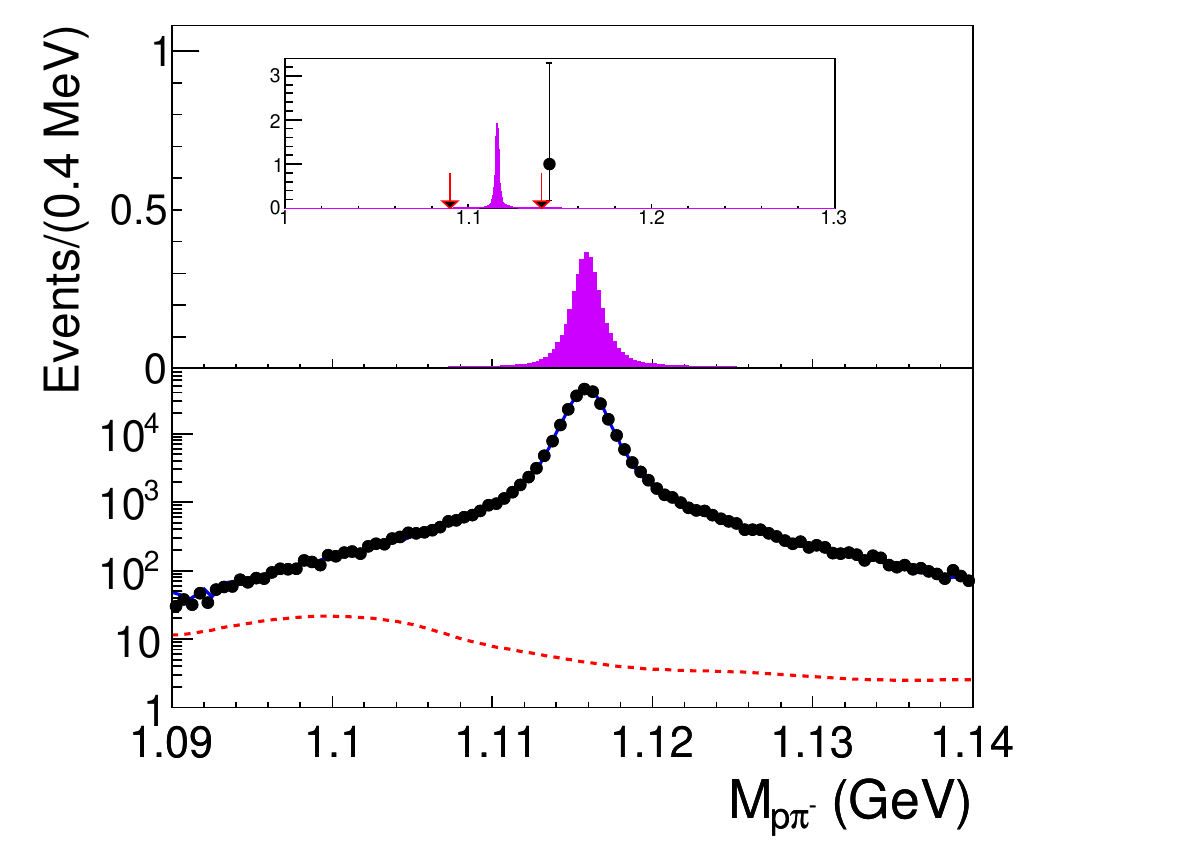}
    \caption{Distribution of $M_{p\pi^-}$ for (a) WS events in the signal region and (insert) over the full span, where the filled circle with error bar is from data, the pink filled histogram, normalized arbitrarily, stems from simulated WS signal events, and the arrows in the inset figure show the edges of the signal region; (b) RS events from data, where the filled circles with error bars are from data, the blue solid line represents the result of the fit and the dashed line shows the background contribution.}
    \label{fig:pklambda}
\end{figure}

The first search for $\Lambda-\bar{\Lambda}$ oscillation has been performed using the decay $J/\psi\to pK^-\bar{\Lambda}$, based on $1.31\times10^{9}$ $J/\psi$ events~\cite{BESIII:2023tge}. Events from the direct decay $J/\psi\to pK^-\bar{\Lambda}$ are designated as $Right$ $Sign$ (RS) events, while those from $J/\psi\to pK^-\bar{\Lambda}\to pK^-\Lambda$ are labeled as $Wrong$ $Sign$ (WS) events. The $\Lambda$ and $\bar{\Lambda}$ baryons are reconstructed via their decays to $p\pi^-$ and $\bar{p}\pi^+$, respectively. Since the $\Lambda$ baryon has a relatively long lifetime, it travels a measurable distance before decaying. A vertex fit is therefore applied by constraining proton and pion tracks to originate from a common decay point. Figure~\ref{fig:pklambda} (b) presents the invariant mass distribution $M_{p\pi^-}$ for the RS sample. A fit to this distribution yields a signal yield of $N^{\mathrm{obs}}_{\mathrm{RS}}=272122\pm528$. In contrast, as shown in Figure~\ref{fig:pklambda} (a), no events survive the WS selection within the signal region of $1.09<M_{p\pi^-}<1.14$ GeV/$c^2$, resulting $N^{\mathrm{obs}}{\mathrm{WS}} = 0$. The detection efficiencies are determined to be $\epsilon_{\mathrm{RS}}=28.6\%$ and $\epsilon_{\mathrm{WS}}=27.8\%$ for the RS and WS processes, respectively.

Assuming no $CP$ violation in $\Lambda$ decay within the process $J/\psi\to pK^-\Lambda$, the $\Lambda-\bar{\Lambda}$ oscillation probability is given by
\begin{equation}
    \mathcal{P}(\Lambda) = \frac{\mathcal{B}(J/\psi\to pK^-\Lambda)}{\mathcal{B}(J/\psi\to pK^-\bar{\Lambda})} = \frac{N^{\mathrm{obs}}_{\mathrm{WS}}/\epsilon_{\mathrm{WS}}}{N^{\mathrm{obs}}_{\mathrm{RS}}/\epsilon_{\mathrm{RS}}}.
\end{equation}
An upper limit (UL) on $\mathcal{P}(\Lambda)$ is set at:
\begin{equation}
    \mathcal{P}(\Lambda)<\frac{s^{90\%}_{\mathrm{WS}}}{N^{\mathrm{obs}}_{\mathrm{RS}}/\epsilon_{\mathrm{RS}}}
    = 4.4\times10^{-6},
\end{equation}
where $s^{90\%}$ denotes the UL at the 90\% confidence level (CL) on the signal events of the WS decay, determined using a frequentist approach~\cite{Rolke:2004mj}. Consequently, the oscillation parameter is constrained to $\delta m_{\Lambda\bar{\Lambda}}<3.8\times10^{-18}$ GeV/$c^2$ at the 90\% CL. This corresponds to a lower limit on the oscillation time, $\tau_{\mathrm{osc}} = 1/\delta m_{\Lambda\bar{\Lambda}}$, of $\tau_{\mathrm{osc}} > 1.7\times10^{-7}$ s at the 90\% CL.

\subsection{Search for $\Lambda-\bar{\Lambda}$ oscillations via $J/\psi\to \Lambda\bar{\Lambda}$}
Using a sample of $(1.0087\pm0.0044)\times10^{10}$ $J/\psi$ decays, a search for $\Lambda-\bar{\Lambda}$ oscillation is performed for the first time via the process $J/\psi\to\Lambda\Lambda+c.c.$~\cite{BESIII:2024gcd}. Events from the decay $J/\psi\to \Lambda\bar{\Lambda}$ are designated as $Right$ $Sign$ (RS) events, while those from $J/\psi\to \Lambda\Lambda+c.c.$ are labeled as $Wrong$ $Sign$ (WS) events. The detection efficiencies are determined to be 34.3\% for the RS channel and 34.4\% for the WS channel. The signal yield for the RS channel is obtained as $3123264\pm1767$. Since no signal or background events are observed in the WS search, the UL on the number of events, $N^{\mathrm{UL}}_{\mathrm{WS}}$, are determined to be 13.0 at the 90\% CL. The $\Lambda-\bar{\Lambda}$ oscillation probability is then calculated as
\begin{equation}
    \mathcal{P}(\Lambda)=\frac{\mathcal{B}(J/\psi\to\Lambda\Lambda+c.c.)}{\mathcal{B}(J/\psi\to\Lambda\bar{\Lambda})} < \frac{N^{\mathrm{UL}}_{\mathrm{WS}}}{N^{\mathrm{obs}}_{\mathrm{RS}}/\epsilon_{\mathrm{RS}}} = 1.4\times10^{-6}.
\end{equation}
Consequently, the UL on the oscillation parameter is set to $\delta m_{\Lambda\bar{\Lambda}} < 2.1 \times 10^{-18}$ GeV/${c^2}$ at the 90\% CL, corresponding to a limit on the oscillation time $\tau_{\mathrm{osc}} > 3.1 \times 10^{-7}$ s. This result establishes a more stringent constraint on $\Lambda-\bar{\Lambda}$ oscillation than the previous measurement from $J/\psi\to pK^-\bar{\Lambda}$ decays.

\section{Search for LNV processes}
\subsection{Study of $D_s^+\to h^+h^0e^+e^-$}
A search for LNV decays $D_s^+\to h^-h^0e^+e^+$ has been performed using 7.33 fb$^{-1}$ of $e^+e^-$ collision data collected at center-of-mass energies from 4.128 to 4.226 GeV~\cite{BESIII:2024ziy}. Here, $h^-$ represents a $K^-$ or $\pi^-$, and $h^0$ represents a $\pi^0$, $K_S^0$ or $\phi$. The analysis covers six decay channels: the Cabibbo-favored $D_s^+\to\phi\pi^-e^+e^+$, the singly Cabibbo-suppressed $D_s^+\to\phi K^-e^+e^+$, $D_s^+\to K_S^0\pi^-e^+e^+$, and the doubly Cabibbo-suppressed $D_s^+\to K_S^0K^-e^+e^+$, $D_s^+\to\pi^-\pi^0e^+e^+$, $D_s^+\to K^-\pi^0e^+e^+$. This work constitutes the first search for four-body $\Delta L=2$ decays of $D_s^+$ meson. In some models~\cite{Atre:2009rg,Yuan:2013yba,Dong:2013raa,Milanes:2018aku}, the predicted branching fractions (BFs) for these processes can reach $\mathcal{O}(10^{-6})$.

\begin{figure}
    \centering
    \includegraphics[width=0.3\linewidth]{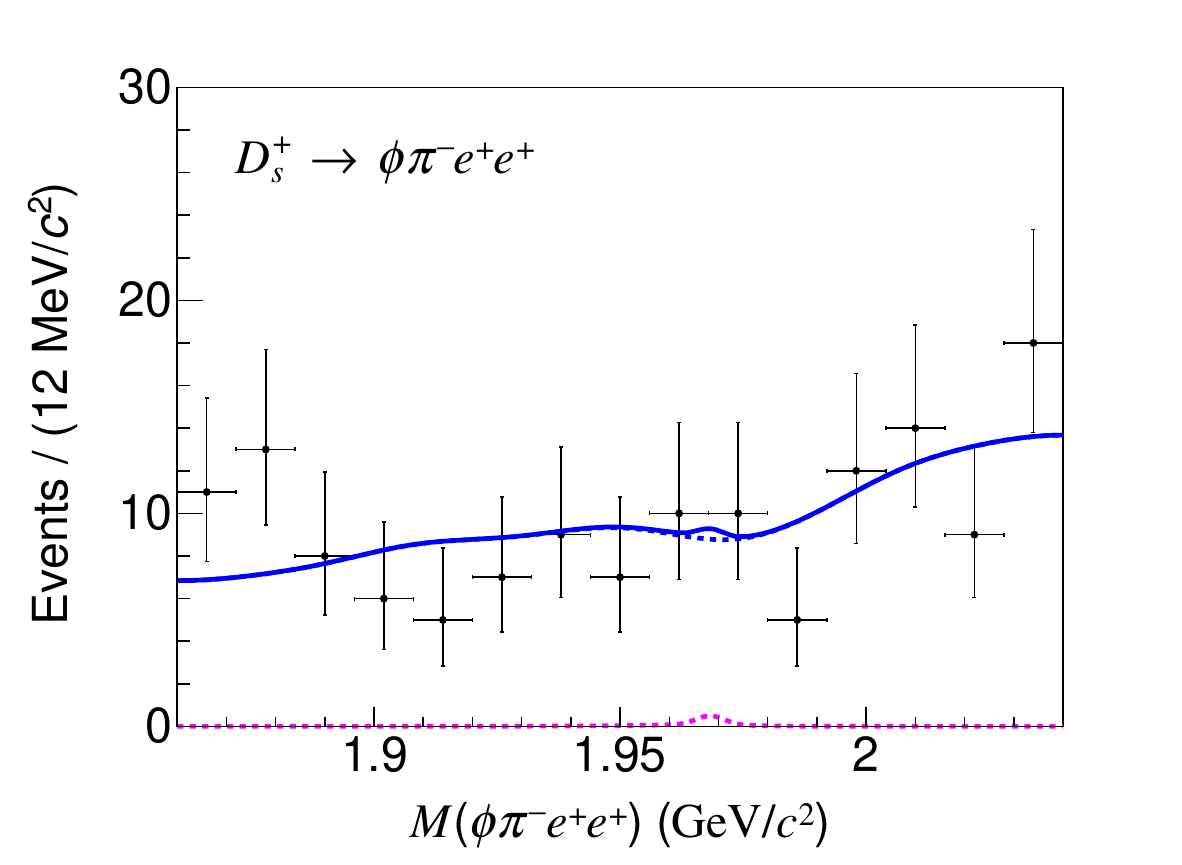}
    \includegraphics[width=0.3\linewidth]{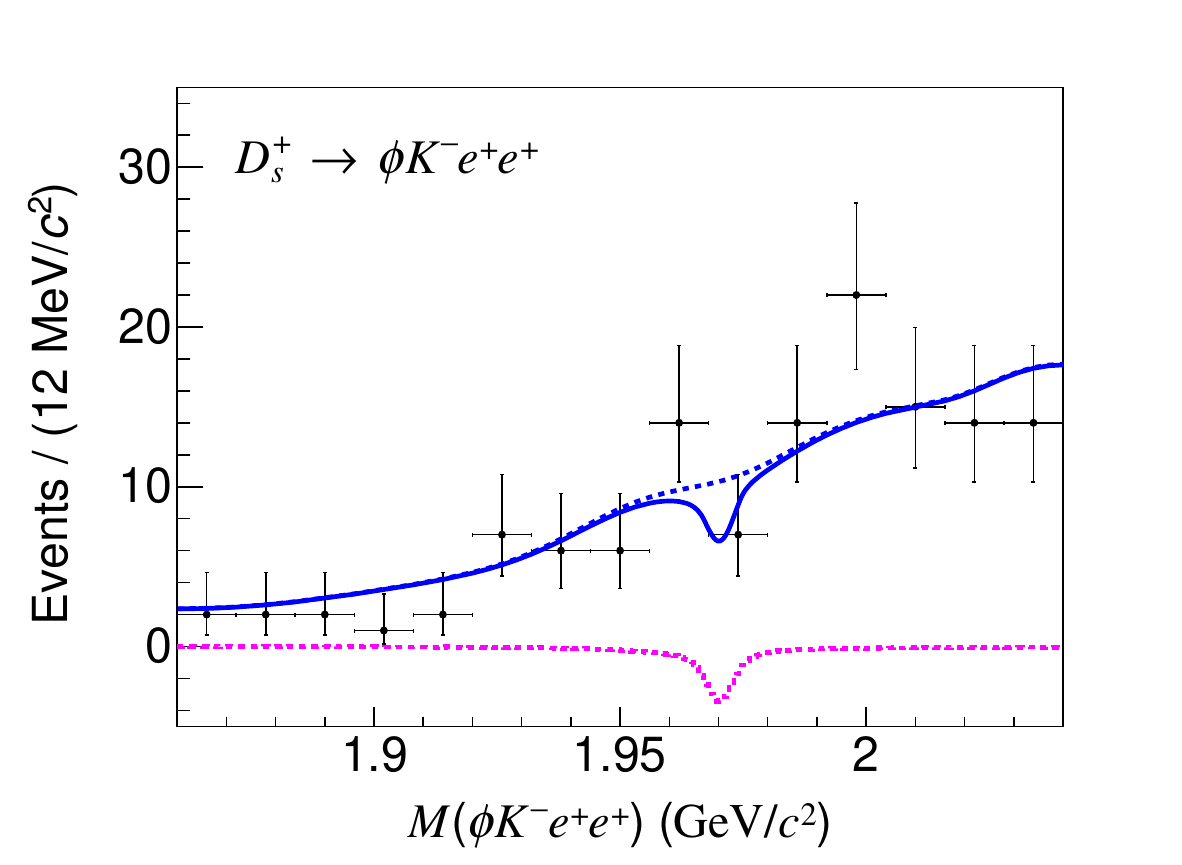}
    \includegraphics[width=0.3\linewidth]{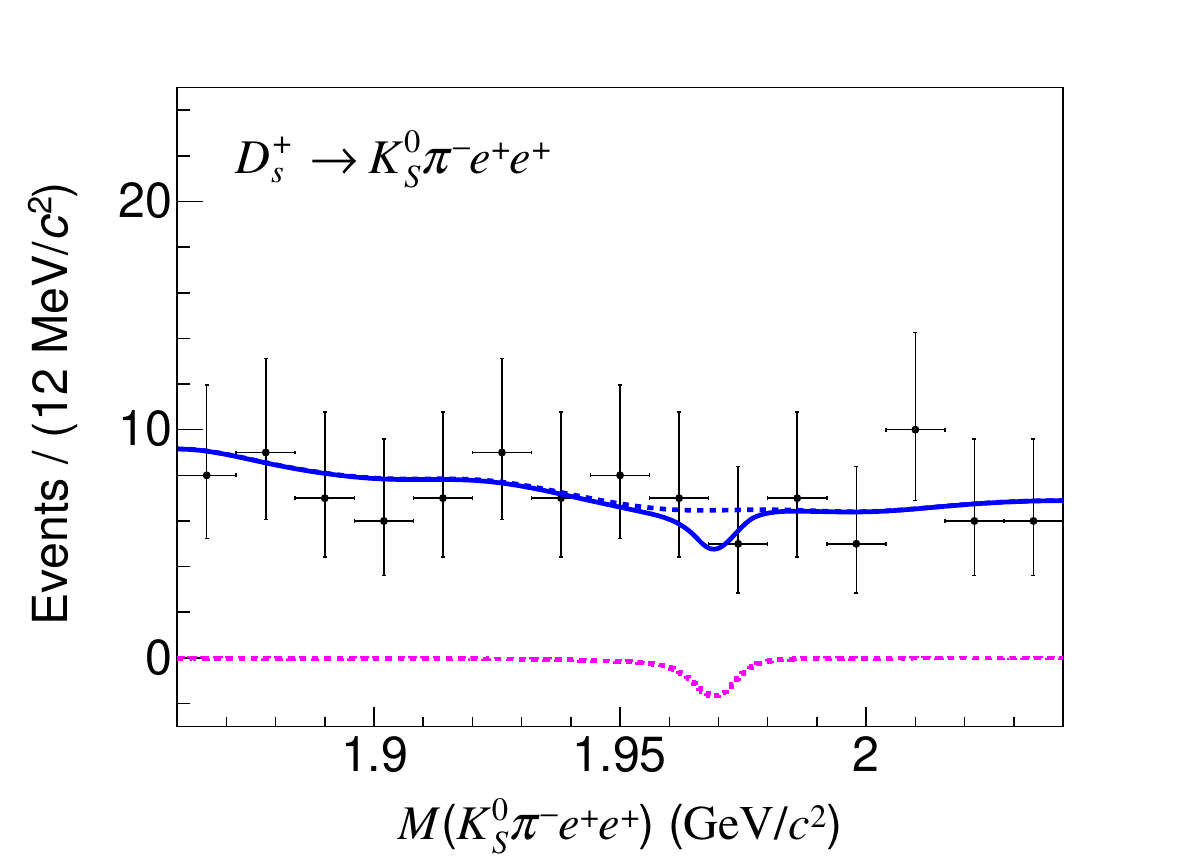}
    \includegraphics[width=0.3\linewidth]{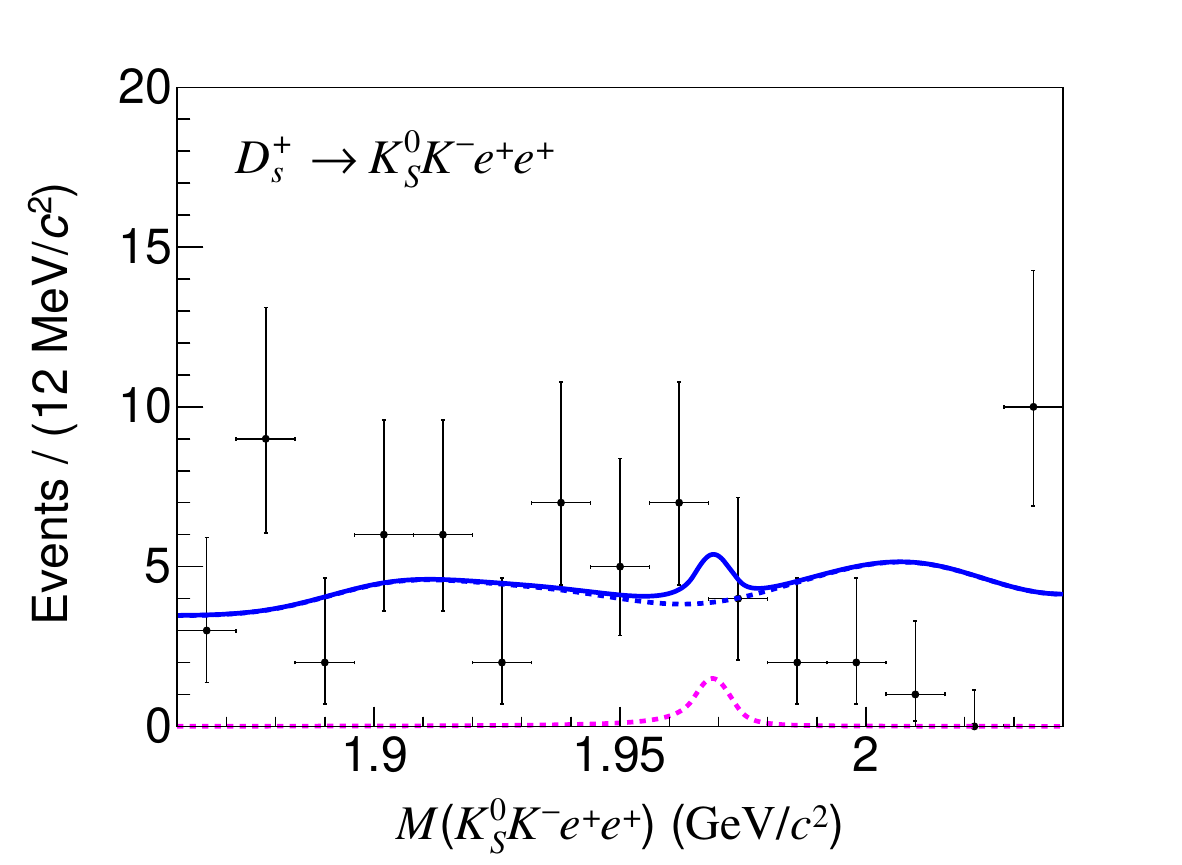}
    \includegraphics[width=0.3\linewidth]{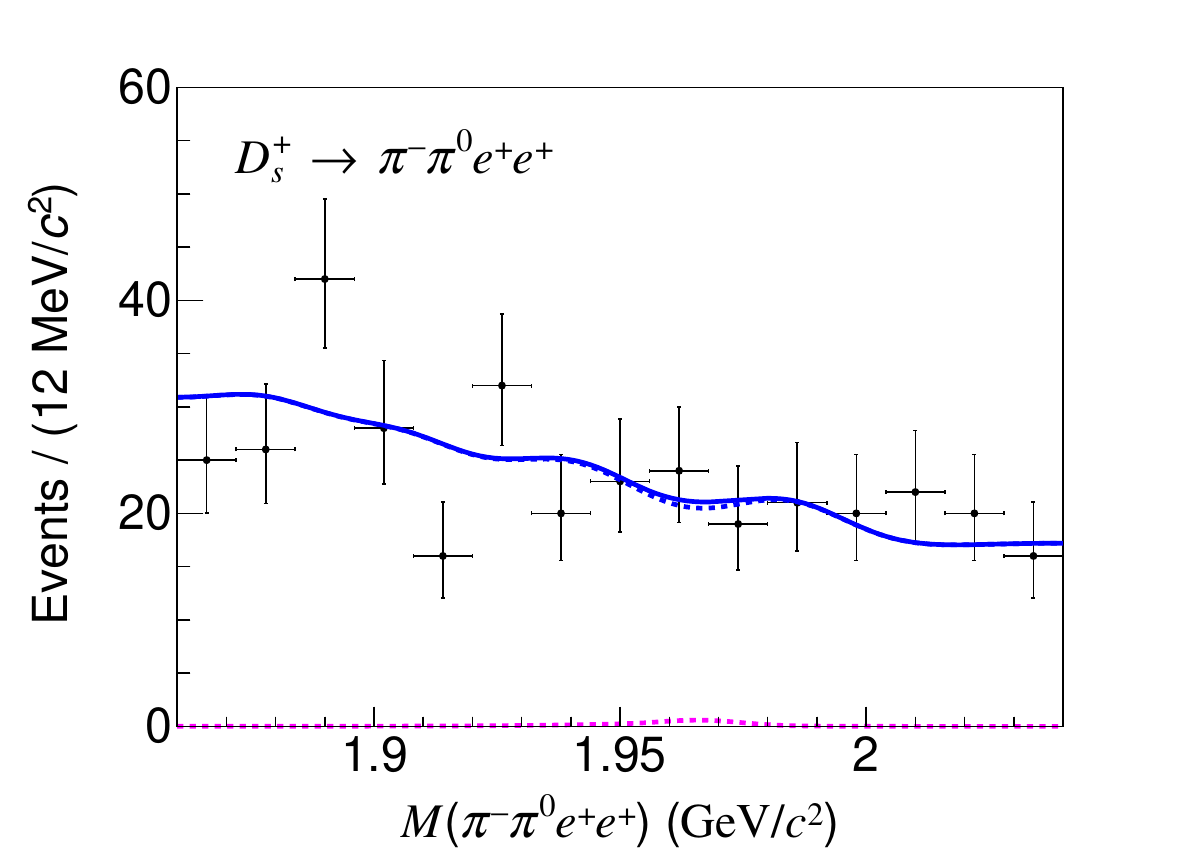}
    \includegraphics[width=0.3\linewidth]{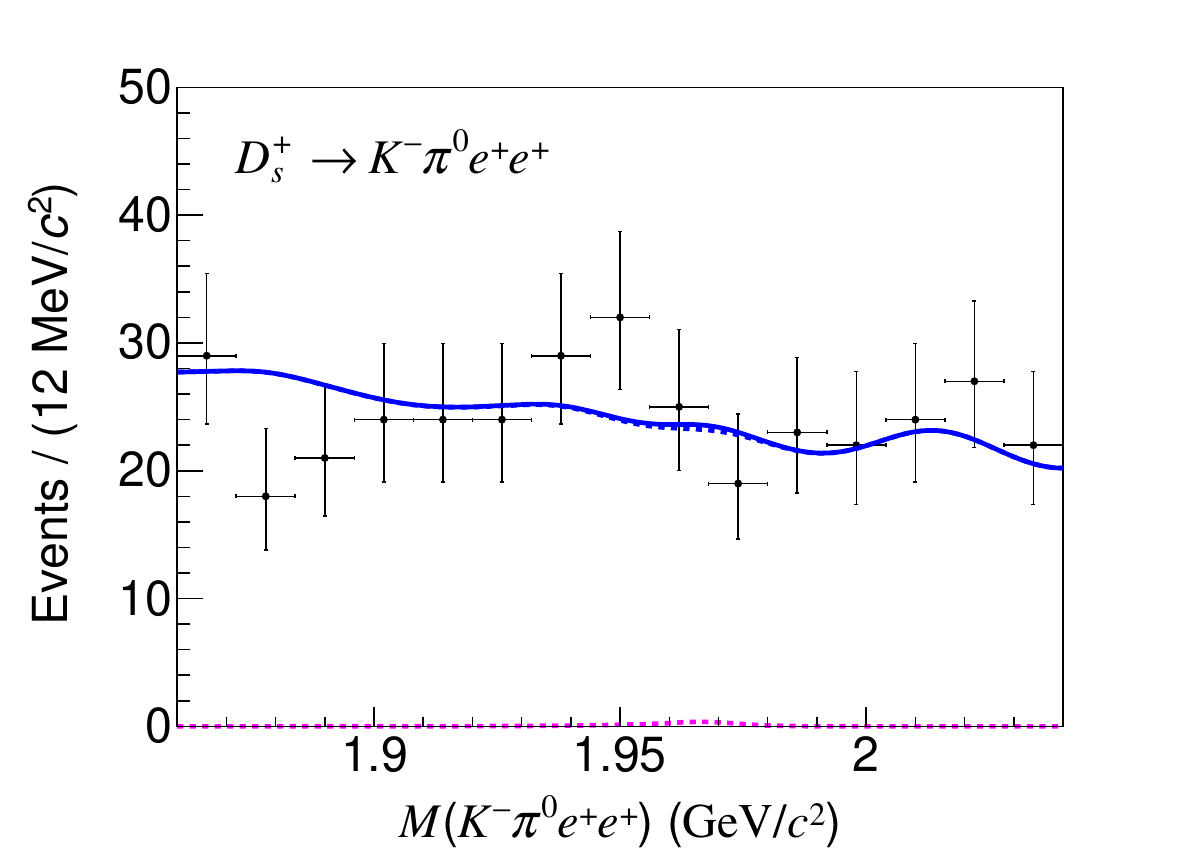}
    \caption{The fits to the invariant mass of corresponding final state of all data samples for each signal channel. The blue solid, magenta dashed and blue dashed lines represent the total fit, signal and background shapes, respectively. The downward magenta peaks denote negative signal yields.}
    \label{fig:hhee_fit}
\end{figure}

In this energy region, $e^+e^-$ annihilate into $D_s^{*\pm}D_s^{\mp}$ pairs. A single-tag method is employed, requiring the reconstruction of only one $D_s^+$ meson per event in the signal decay mode. The BF is calculated by
\begin{equation}
    \mathcal{B}(D_s^+\to h^-h^0e^+e^+) = \frac{N_{\mathrm{sig}}}{2\cdot N_{D_s^{*\pm}D_s^{\mp}}\cdot\epsilon\cdot\mathcal{B}_{\mathrm{inter}}},
\end{equation}
where $N_{\mathrm{sig}}$ is the signal yield, $N_{D_s^{*\pm}D_s^{\mp}}$ is the total number of $D_s^{*\pm}D_s^{\mp}$ pairs, and $\epsilon$ is the detection efficiency, and $\mathcal{B}_{\mathrm{inter}}$ represents the product of intermediate BFs. The signal is identified in the invariant mass spectrum, as shown in Figure~\ref{fig:hhee_fit}. Since no obvious signal is observed, the ULs on the BFs of $D_s^+\to h^-h^0e^+e^+$ decays are set using a Bayesian method~\cite{Zhu:2007zza}. Based on a background-only hypothesis, expected ULs of the BFs are also determined. Both expected and observed ULs of the BFs at the 90\% CL are summarized in Table~\ref{tab:hhee_BUL}.

\begin{figure}
    \centering
    \includegraphics[width=0.8\linewidth]{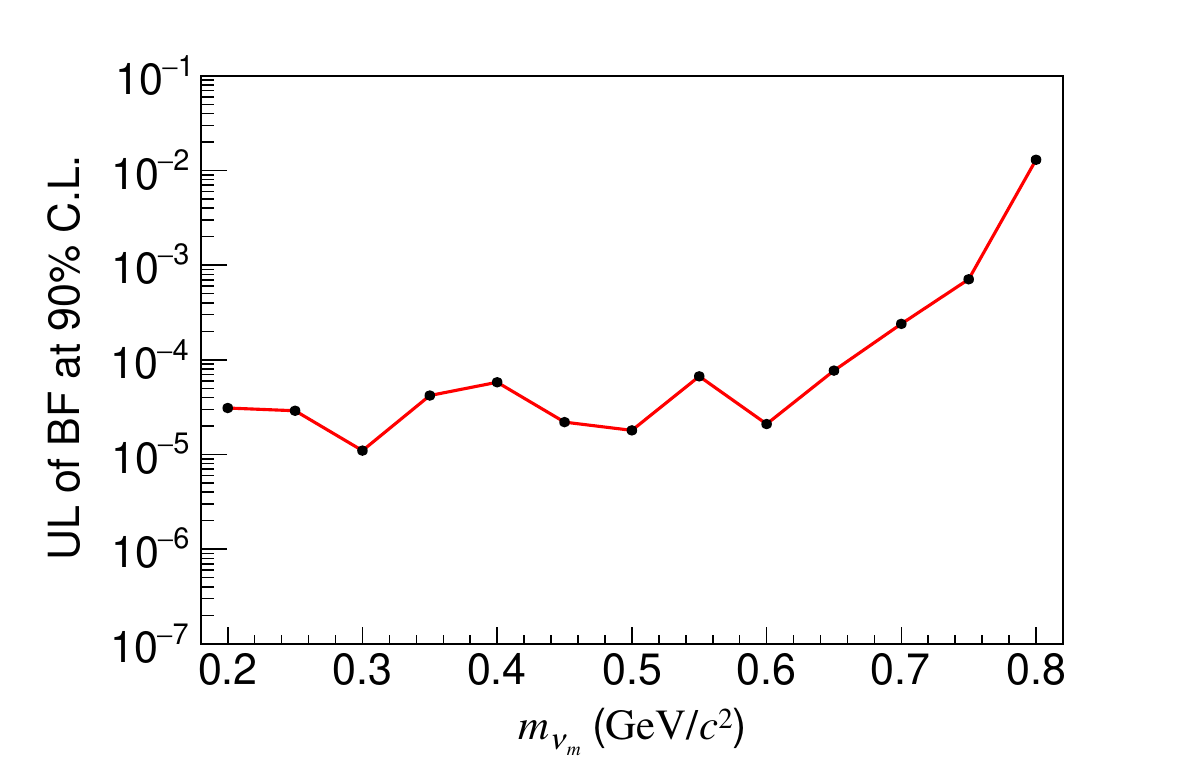}
    \caption{The ULs of the BFs at the 90\% CL as a function of $m_{\nu_{m}}$ for the $D_s^+ \to \phi e^+ \nu_{m}(\to\pi^-e^+)$ decay.}
    \label{fig:mvm}
\end{figure}

Additionally, the Majorana neutrino is searched for in the decay of $D_s^+\to\phi e^+ \nu_m (\to \pi^- e^+)$ with various mass $m_{\nu_{m}}$ assumptions, ranging from 0.20 to 0.80 GeV/$c^2$ in intervals of 0.05 GeV/$c^2$. For each assumed mass $m_{\nu_{m}}$, candidate events are selected by requiring the invariant mass $M_{\pi^-e^+}$ of any $\pi^-e^+$ combination to be within the range of $[m_{\nu_m}-5\sigma,m_{\nu_m}+4\sigma]$, where $\sigma$ is the resolution of the $M_{\pi^-e^+}$ distribution obtained by MC simulation. Using a profile likelihood method~\cite{Rolke:2004mj}, the ULs on the BFs at the 90\% CL are determined as a function of $m_{\nu_{m}}$, ranging from around $10^{-5}-10^{-2}$, as shown in Figure~\ref{fig:mvm}.

\begin{table}[htbp]
    \centering
    \begin{tabular}{cc}
        \hline
       Decay channel & $\mathcal{B}_{\mathrm{UL}}(\mathcal{B}^{\mathrm{expected}}_{\mathrm{UL}})$ \\
       \hline
       $D_s^+\to\phi\pi^-e^+e^+$  & $6.9 (3.5)\times10^{-5}$ \\
       $D_s^+\to\phi K^-e^+e^+$  & $9.9 (10.8)\times10^{-5}$ \\
       $D_s^+\to K_S^0\pi^-e^+e^+$  & $1.3 (2.4)\times10^{-5}$ \\
       $D_s^+\to K_S^0K^-e^+e^+$  & $2.9 (2.3)\times10^{-5}$ \\
       $D_s^+\to \pi^{-}\pi^{0}e^+e^+$  & $2.9 (2.7)\times10^{-5}$ \\
       $D_s^+\to K^-\pi^0e^+e^+$  & $3.4 (3.9)\times10^{-5}$ \\
       \hline
    \end{tabular}
    \caption{The observed and expected ULs of the BFs at the 90\% CL ($\mathcal{B}_{\mathrm{UL}}$ and $\mathcal{B}^{\mathrm{expected}}_{\mathrm{UL}}$) of $D_s^+\to h^-h^0e^+e^+$ decays.}
    \label{tab:hhee_BUL}
\end{table}

\subsection{Study of $\omega\to \pi^+\pi^+e^-e^-$}
\begin{figure}
    \centering
    \includegraphics[width=0.4\linewidth]{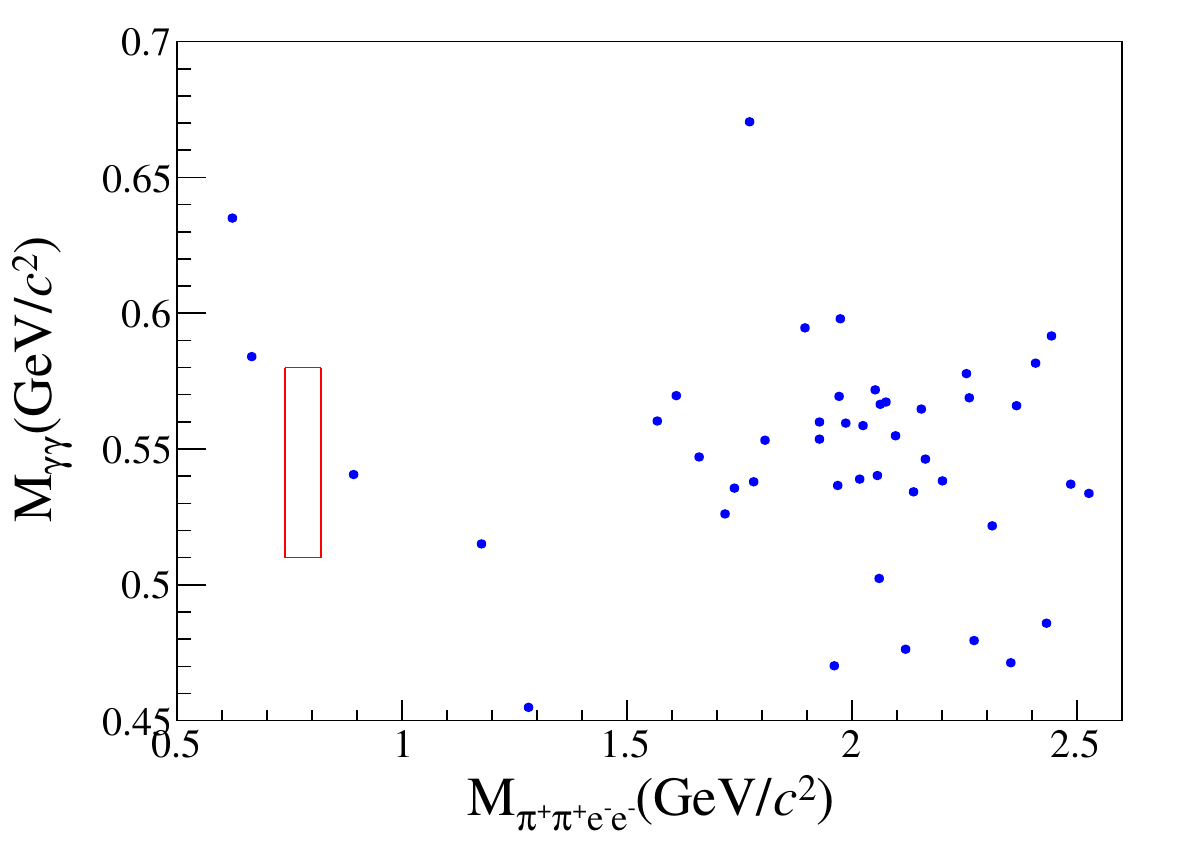}
    \includegraphics[width=0.42\linewidth]{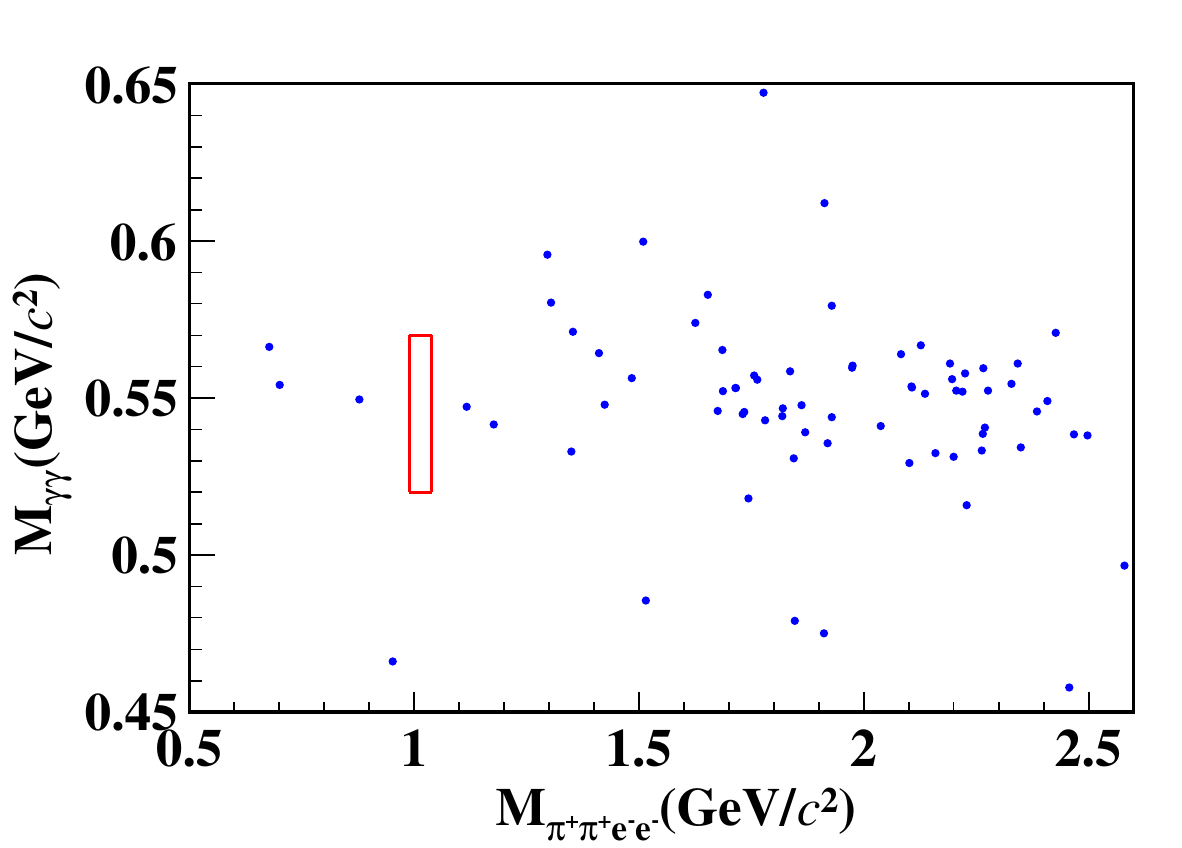}
    \caption{The 2D distribution of $M_{\gamma\gamma}$ versus $M_{\pi^+\pi^+e^-e^-}$ of the candidate events from the $J/\psi$ data, where the red box shows the signal region. The left plot is for $\omega\to\pi^+\pi^+e^-e^-$, and the right plot is for $\phi\to\pi^+\pi^+e^-e^-$.}
    \label{fig:pipiee}
\end{figure}

LNV decay $\omega\to\pi^+\pi^+e^-e^-$ is searched for via $J/\psi\to\eta\omega$ using a data sample of $(1.0087\pm0.0044)\times10^{10}$ $J/\psi$ events~\cite{BESIII:2025gsy}. The measurement is performed relative to the normalization channel $\omega\to\pi^+\pi^-\pi^0$. As shown in Figure~\ref{fig:pipiee}, no candidates are found within the signal region. Using Feldman-Cousins intervals~\cite{Feldman:1997qc, Rolke:2004mj}, the UL on the BF at the 90\% CL is set to be $\mathcal{B}(\omega\to\pi^+\pi^+e^-e^-)<2.8\times10^{6}$.

\subsection{Study of $\phi\to \pi^+\pi^-e^-e^-$}
LNV decay $\phi\to\pi^+\pi^+e^-e^-$ is searched for via $J/\psi\to\eta\phi$ using a data sample of $(1.0087\pm0.0044)\times10^{10}$ $J/\psi$ events~\cite{BESIII:2023pqp}. The measurement is performed relative to the normalization channel $\phi\to K^+K^-$. As shown in Figure~\ref{fig:pipiee}, no candidates are found within the signal region. Using a frequentist method, the UL on the BF at the 90\% CL is set to be $\mathcal{B}(\phi\to\pi^+\pi^+e^-e^-)<3.7\times10^{-6}$. This work establishes the first constraint on LNV in $\phi$ meson decays, thereby improving the experimental knowledge of neutrinoless double beta decay mechanisms in hadrons composed of second-generation quarks.

\section{Search for BNV\&LNV processes}
\subsection{Study of $\Xi^0\to K^{\mp}e^{\pm}$}
\begin{figure}
    \centering
    \includegraphics[width=0.4\linewidth]{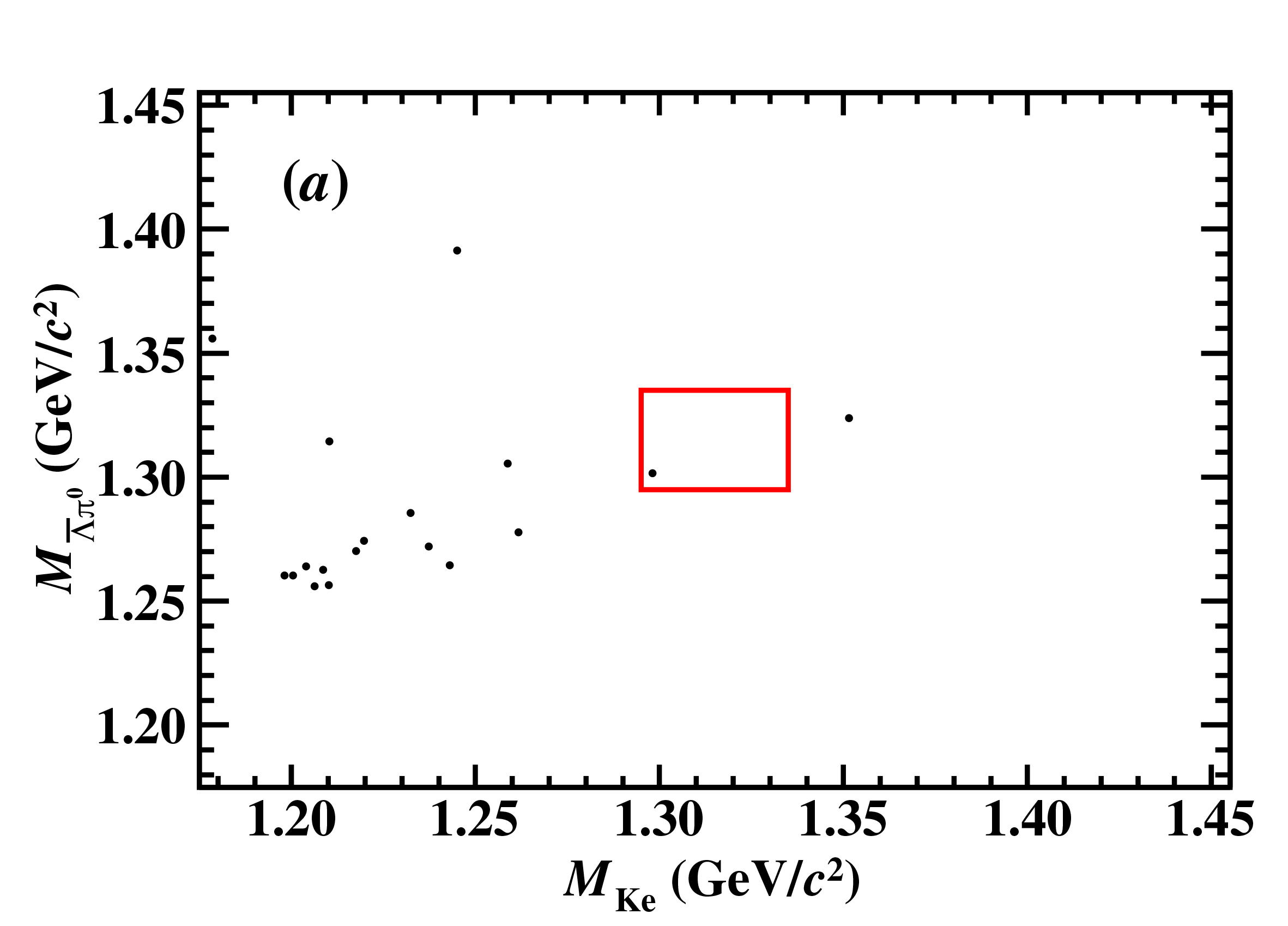}
    \includegraphics[width=0.42\linewidth]{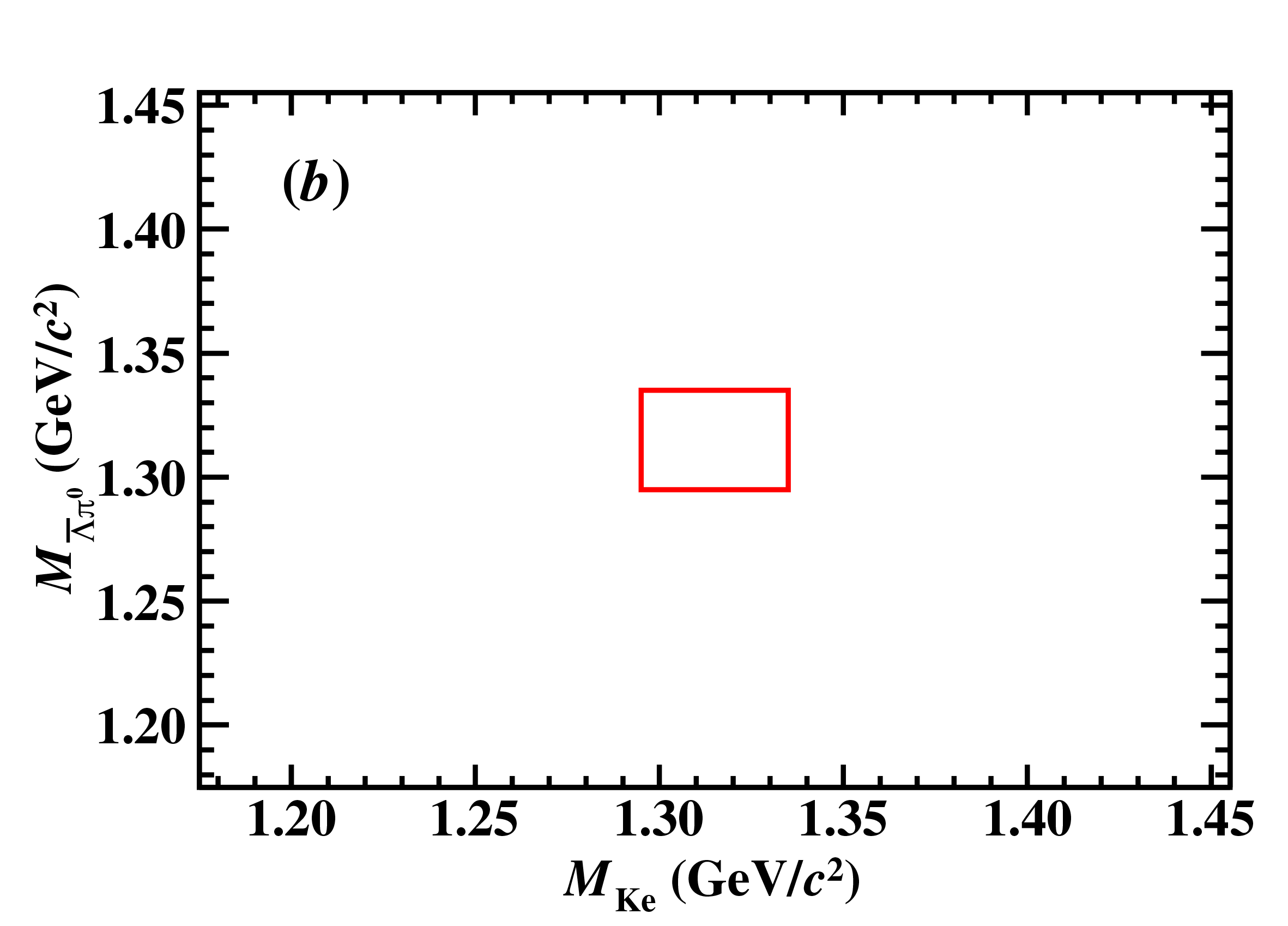}
    \caption{Distributions of $M_{Ke}$ versus $M_{\bar{\Lambda}\pi^0}$ of the accepted candidates for (a) $\Xi^0\to K^-e^+$ and (b) $\Xi^0\to K^+e^-$ in data, respectively. The red box indicates the signal region.}
    \label{fig:ke}
\end{figure}

The first search for the simultaneous violation of baryon and lepton numbers in $\Xi^0$ decays has been conducted, analyzing $\Xi^0\to K^-e^+$ with $\Delta (B-L)=0$ and $\Xi^0\to K^+e^-$ with $|\Delta (B-L)|=2$ using $(1.0087\pm0.0044)\times10^{10}$ $J/\psi$ events~\cite{BESIII:2023str}. The $\Xi^0\bar{\Xi}^0$ hyperon pairs are produced in $J/\psi$ decays without any additional fragmentation particles. Double tag method is implemented, wherein one $\bar{\Xi}^0$ hyperon is fully reconstructed through the decay mode $\bar{\Xi}^0\to\bar{\Lambda}(\to\bar{p}\pi^+)\pi^0(\to\gamma\gamma)$, and the accompanying $\Xi^0$ is reconstructed in the signal decay. As shown in Figure~\ref{fig:ke}, one event is observed in the signal region of $\Xi^0\to K^-e^+$, while no event is observed in the signal region of $\Xi^0\to K^+e^-$. Using a frequentist method of profile likelihood~\cite{Rolke:2004mj}, the ULs on the BFs at the 90\% CL are set to be $\mathcal{B}(\Xi^0\to K^-e^+)<3.6\times10^{-6}$ and $\mathcal{B}(\Xi^0\to K^+e^-)<1.9\times10^{-6}$, among the best constraints on the BNV interactions from hyperon decays.

\section{Summary}
We present new results from the BESIII experiment on searches for BNV and LNV processes. This includes the first study of $\Lambda-\bar{\Lambda}$ oscillations in $J/\psi\to pK^-\bar{\Lambda}$, the most stringent constraints on $\Lambda-\bar{\Lambda}$ oscillation parameters from $J/\psi\to\Lambda\bar{\Lambda}$, the first investigation of four-body LNV decays $D_s^+\to h^-h^0e^+e^+$, searches for light flavor meson $\omega/\phi$ LNV decays into $\pi^+\pi^+e^-e^-$, and $\Xi^0\to K^{\mp}e^{\pm}$ decays. These results provide significant insights into the matter-antimatter asymmetry of the universe and probe for new physics beyond Standard Model.


\begin{thebibliography}{99}

\bibitem{Georgi:1974sy}
H.~Georgi and S.~L.~Glashow,
Phys. Rev. Lett. \textbf{32} (1974), 438-441
doi:10.1103/PhysRevLett.32.438

\bibitem{Milanes:2018aku}
D.~Milan{\'e}s and N.~Quintero,
Phys. Rev. D \textbf{98} (2018) no.9, 096004
doi:10.1103/PhysRevD.98.096004
[arXiv:1808.06017 [hep-ph]].

\bibitem{Li:2024moj}
Z.~Li and Z.~You,
[arXiv:2403.11597 [hep-ex]].

\bibitem{Sakharov:1967dj}
A.~D.~Sakharov,
Pisma Zh. Eksp. Teor. Fiz. \textbf{5} (1967), 32-35
doi:10.1070/PU1991v034n05ABEH002497

\bibitem{Babu:2012iv}
K.~S.~Babu and R.~N.~Mohapatra,
Phys. Rev. Lett. \textbf{109} (2012), 091803
doi:10.1103/PhysRevLett.109.091803
[arXiv:1207.5771 [hep-ph]].

\bibitem{BESIII:2009fln}
M.~Ablikim \textit{et al.} [BESIII Collaboration],
Nucl. Instrum. Meth. A \textbf{614} (2010), 345-399
doi:10.1016/j.nima.2009.12.050
[arXiv:0911.4960 [physics.ins-det]].

\bibitem{Li:2024pox}
Z.~J.~Li, M.~K.~Yuan, Y.~X.~Song, Y.~G.~Li, J.~S.~Li, S.~S.~Sun, X.~L.~Wang, Z.~Y.~You and Y.~J.~Mao,
Front. Phys. (Beijing) \textbf{19} (2024) no.6, 64201
doi:10.1007/s11467-024-1422-7
[arXiv:2404.07951 [physics.data-an]].

\bibitem{Song:2025pnt}
T.~Z.~Song, K.~X.~Huang, Y.~J.~Zeng, M.~H.~Liao, X.~S.~Wang, Y.~M.~Zhang and Z.~Y.~You,
Front. Phys. (Beijing) \textbf{21} (2026) no.2, 26201
doi:10.15302/frontphys.2026.026201
[arXiv:2507.10261 [hep-ex]].

\bibitem{Liao:2025lth}
M.~H.~Liao, J.~S.~Liu, X.~N.~Wang, S.~S.~Sun and Z.~Y.~You,
Nucl. Sci. Tech. \textbf{36} (2025) no.11, 218
doi:10.1007/s41365-025-01789-y
[arXiv:2509.16066 [hep-ex]].

\bibitem{Mohapatra:1980qe}
R.~N.~Mohapatra and R.~E.~Marshak,
Phys. Rev. Lett. \textbf{44} (1980), 1316-1319
[erratum: Phys. Rev. Lett. \textbf{44} (1980), 1643]
doi:10.1103/PhysRevLett.44.1316

\bibitem{Kuzmin:1970nx}
V.~A.~Kuzmin,
Pisma Zh. Eksp. Teor. Fiz. \textbf{12} (1970) no.6, 335-337

\bibitem{Baldo-Ceolin:1994hzw}
M.~Baldo-Ceolin, P.~Benetti, T.~Bitter, F.~Bobisut, E.~Calligarich, R.~Dolfini, D.~Dubbers, P.~El-Muzeini, M.~Genoni and D.~Gibin, \textit{et al.}
Z. Phys. C \textbf{63} (1994), 409-416
doi:10.1007/BF01580321

\bibitem{Kang:2009xt}
X.~W.~Kang, H.~B.~Li and G.~R.~Lu,
Phys. Rev. D \textbf{81} (2010), 051901
doi:10.1103/PhysRevD.81.051901
[arXiv:0906.0230 [hep-ph]].

\bibitem{BESIII:2023tge}
M.~Ablikim \textit{et al.} [BESIII Collaboration],
Phys. Rev. Lett. \textbf{131} (2023) no.12, 12
doi:10.1103/PhysRevLett.131.121801
[arXiv:2305.04568 [hep-ex]].

\bibitem{Rolke:2004mj}
W.~A.~Rolke, A.~M.~Lopez and J.~Conrad,
Nucl. Instrum. Meth. A \textbf{551} (2005), 493-503
doi:10.1016/j.nima.2005.05.068
[arXiv:physics/0403059 [physics]].

\bibitem{BESIII:2024gcd}
M.~Ablikim \textit{et al.} [BESIII Collaboration],
Phys. Rev. D \textbf{111} (2025) no.5, 052014
doi:10.1103/PhysRevD.111.052014
[arXiv:2410.21841 [hep-ex]].

\bibitem{BESIII:2024ziy}
M.~Ablikim \textit{et al.} [BESIII Collaboration],
JHEP \textbf{01} (2025), 109
doi:10.1007/JHEP01(2025)109
[arXiv:2410.02421 [hep-ex]].

\bibitem{Atre:2009rg}
A.~Atre, T.~Han, S.~Pascoli and B.~Zhang,
JHEP \textbf{05} (2009), 030
doi:10.1088/1126-6708/2009/05/030
[arXiv:0901.3589 [hep-ph]].

\bibitem{Yuan:2013yba}
H.~Yuan, T.~Wang, G.~L.~Wang, W.~L.~Ju and J.~M.~Zhang,
JHEP \textbf{08} (2013), 066
doi:10.1007/JHEP08(2013)066
[arXiv:1304.3810 [hep-ph]].

\bibitem{Dong:2013raa}
H.~R.~Dong, F.~Feng and H.~B.~Li,
Chin. Phys. C \textbf{39} (2015) no.1, 013101
doi:10.1088/1674-1137/39/1/013101
[arXiv:1305.3820 [hep-ph]].

\bibitem{Zhu:2007zza}
Y.~Zhu,
Nucl. Instrum. Meth. A \textbf{578} (2007), 322-328
doi:10.1016/j.nima.2007.05.116

\bibitem{Rolke:2004mj}
W.~A.~Rolke, A.~M.~Lopez and J.~Conrad,
Nucl. Instrum. Meth. A \textbf{551} (2005), 493-503
doi:10.1016/j.nima.2005.05.068
[arXiv:physics/0403059 [physics]].

\bibitem{BESIII:2025gsy}
M.~Ablikim \textit{et al.} [BESIII Collaboration],
[arXiv:2504.21539 [hep-ex]].

\bibitem{Feldman:1997qc}
G.~J.~Feldman and R.~D.~Cousins,
Phys. Rev. D \textbf{57} (1998), 3873-3889
doi:10.1103/PhysRevD.57.3873
[arXiv:physics/9711021 [physics.data-an]].

\bibitem{BESIII:2023pqp}
M.~Ablikim \textit{et al.} [BESIII Collaboration],
Chin. Phys. C \textbf{49} (2025) no.4, 043001
doi:10.1088/1674-1137/ada350
[arXiv:2308.05490 [hep-ex]].

\bibitem{BESIII:2023str}
M.~Ablikim \textit{et al.} [BESIII Collaboration],
Phys. Rev. D \textbf{108} (2023) no.1, 012006
doi:10.1103/PhysRevD.108.012006
[arXiv:2305.07231 [hep-ex]].

\end{thebibliography}
\end{document}